\documentclass[aps,prb,floatfix,twocolumn,superscriptaddress,showpacs,amsmath,amssymb]{revtex4}
\usepackage{graphicx}
\usepackage{epstopdf}
\usepackage{epsfig} 
\usepackage{mathrsfs}
\usepackage{color}
\allowdisplaybreaks
\newcommand{\be}{\begin{equation}}
\newcommand{\ee}{\end{equation}}
\newcommand{\bea}{\begin{eqnarray}}
\newcommand{\eea}{\end{eqnarray}}
\newcommand{\ba}{\begin{eqnarray*}}
\newcommand{\ea}{\end{eqnarray*}}
\newenvironment{eqs}%
{\begin{equation} \begin{aligned}}%
{\end{aligned} \end{equation} }
\newcommand{\bal}{\begin{eqs}}
\newcommand{\eal}{\end{eqs}}
{\begin{equation} \begin{split}}%
{\end{split} \end{equation} }
\newcommand{\bas}{\begin{eqs}}
\newcommand{\eas}{\end{eqs}}
\newcommand{\dagga}{{\phantom{\dagger}}}

\newcommand{\bk}{\mathbf{k}}

\newcommand{\bp}{\mathbf{p}}

\newcommand{\bC}[2]{\mathbf{C}_{#1}(#2)}
\newcommand{\dis}{\displaystyle}
\newcommand{\up}{\uparrow}
\newcommand{\down}{\downarrow}
\newcommand{\fract}[2]{\frac{\dis #1}{\dis #2}}
\newcommand{\Tr}{\text{Tr}}
\newcommand{\eqn}[1]{(\ref{#1})}
\newcommand{\ket}[1]{\mid\! #1\rangle}
\newcommand{\bra}[1]{\langle #1\!\mid} 
\newcommand{\bw}{\begin{widetext}}
\newcommand{\ew}{\end{widetext}}
\newcommand{\esp}[1]{\text{e}^{#1}}

\newcommand{\ep}{{\epsilon}}

\begin{document}

\title{Manumitting slave-spins in the Anderson impurity model}

\author{Daniele Guerci} 
\affiliation{International School for
  Advanced Studies (SISSA), Via Bonomea
  265, I-34136 Trieste, Italy}  
\author{Michele Fabrizio} 
\affiliation{International School for
  Advanced Studies (SISSA), Via Bonomea
  265, I-34136 Trieste, Italy} 

\date{\today} 


\begin{abstract}
We show that a generic single-orbital Anderson impurity model, lacking for instance any kind of particle-hole symmetry, can be exactly mapped without any constraint onto a resonant level model coupled to two Ising variables, which reduce to one if the hybridisation is particle-hole symmetric. 
The mean-field solution of this model is found to be stable to unphysical spontaneous magnetisation of the impurity, unlike the saddle-point solution in the standard slave-boson representation. Remarkably, the mean-field estimate of the Wilson ratio approaches the exact value $R_\text{W}=2$ in the Kondo regime. 
\end{abstract}
\maketitle

\section{Introduction}

Within any approximate technique based on independent particles, 
as e.g. Hartree-Fock, the electron's quantum numbers, i.e. its charge, spin and, eventually, orbital component, 
are inevitably all entangled into single-particle excitations. This is ultimately the reason why such 
independent-particle schemes fail in correlated electron systems 
where charge degrees of freedom are instead well separated 
in energy from spin and orbital ones.\\
An efficient and popular trick to disentangle charge from other degrees of freedom is to enlarge the Hilbert space adding auxiliary particles \textit{slaves} to the physical charge excitations. There are by now various implementations of such trick, starting from the elder slave-boson theory~\cite{Barnes-slave-boson,Coleman-slave-boson,Kotliar&Ruckenstein} to more recent 
slave-spin~\cite{De-Medici,Z2-1,Z2-2} and slave-rotor~\cite{slave-rotors} ones. Those auxiliary particles are held in slavery by a product of local constraints that project the enlarged Hilbert space $\mathscr{H}_*$
onto the physical subspace $\mathscr{H}$, and concurrently the effective Hamiltonian $H_*$ of the electrons plus the auxiliary particles onto the original electron-only one, $H$. As common in such cases, 
$H_*$ possesses local gauge invariance that translates into local conserved quantities. The constraints simply fix the values that those conserved quantities must have in the physical subspace. \\
The big advantage of this apparently more cumbersome approach is that a mean-field decoupling of the electrons from the slave particles naturally provides the desired disentanglement of charge from all other degrees of freedom, thus allowing the access to phenomena like Mott's localisation~\cite{Kotliar&Ruckenstein} otherwise inaccessible by mean-field in the original 
electron-only representation.\\
The problem with mean field in slave-particle theories is that the constraints are only satisfied \textit{on average}, which brings about unphysical gauge-symmetry breaking, i.e. mean-field solutions mixing the physical subspace with the non-physical one. 
There is actually an exception where the constraint is not required:  a particle-hole (p-h) symmetric single-orbital Anderson impurity model (AIM) that is represented in terms of a resonant level coupled to a two-level system, one level corresponding to the impurity being singly occupied and the other to the impurity being empty or doubly occupied. Because of p-h symmetry, the partition function within the physical subspace 
is equal~\cite{Marco-PRB} to that in the unphysical one, so that the former is just  
half of the partition function calculated in the whole enlarged Hilbert space without any restriction. In this representation the Hamiltonian possesses a local $Z_2$ gauge symmetry, which is spontaneously broken at zero temperature~\cite{Pierpaolo} since the model effectively corresponds to a two-level system in a sub-ohmic bath~\cite{Caldeira&Leggett}. Therefore the symmetry breaking is here not a spurious result of mean field but a real feature of the model. Since a p-h symmetric Hubbard model in 
infinitely coordinated lattices maps within dynamical mean-field theory (DMFT)~\cite{DMFT} just onto that same AIM, one can show~\cite{Marco-PRB} that the free energy of the lattice model can be straightforwardly obtained by that of its $Z_2$ slave spin representation~\cite{Z2-1,Z2-2} without imposing any constraint. One remarkable consequence of such mapping is that the metallic phase of the Hubbard model translates into a phase where the local $Z_2$ gauge symmetry 
breaks spontaneously~\cite{Rok-PRB2015}, which is not prohibited when the lattice coordination number is infinite~\cite{Maslanka}, whereas the symmetry is restored in the Mott insulator. This mapping thus endows the Mott transition of a genuine order parameter. More recently, a similar trick of exploiting particle-hole symmetry to get rid of the local constraints was used~\cite{Gabi-slave-spin} to derive a Landau-Ginzburg theory of the orbital-selective Mott transition in a two-band Hubbard model at half-filling. \\
In view of the above promising results, it is worth exploring whether it is still possible to get rid of the constraints away from  particle-hole symmetry, which is precisely the goal of the present work.  

\section{The model}
\label{The model}
We consider the single-orbital AIM
\be
\begin{split}
H&= \sum_{\bk\sigma}\,\Bigg[\ep_{\bk\sigma}\,c^\dagger_{\bk\sigma}\,c^\dagga_{\bk\sigma} 
+ T_{\bk\sigma}\Big(d^\dagger_\sigma\,c^\dagga_{\bk\sigma} + c^\dagger_{\bk\sigma}\,d^\dagga_\sigma\Big)\Bigg]
\\
&\qquad -\fract{U}{4}\,\Omega -\mu\big(n_\up+n_\down-1\big) - h\big(n_\up-n_\down\big)\,,
\end{split}\label{HAIM}
\ee
where $n_\sigma = d^\dagger_\sigma d^\dagga_\sigma$ and 
\be
\Omega=\Omega^\dagger=\Omega^{-1}= -\big(2n_\up-1\big)\big(2n_\down-1\big)\,,
\label{Omega}
\ee
such that $\Omega\,d^\dagga_\sigma\,\Omega = - d^\dagga_\sigma$. We assume generically spin-dependent and p-h non-symmetric hybridisation amplitudes $T_{\bk\sigma}$. By contrast, we can always consider, without loss of generality, a p-h 
symmetric spectrum $\ep_{\bk\sigma}$, which implies the existence of 
a one-to-one correspondence between spin-dependent pairs of momenta, $\bk$ and $\bp=\bC{\sigma}{\bk}$, such that $\ep_{\bk\sigma}=-\ep_{\bp\sigma}$. For convenience we define for all $\bk$ such 
that $\ep_\bk<0$ the following combinations of fermionic operators 
\bal
c^\dagga_{1(2)\bk\sigma} = 
\Big(c^\dagga_{\bk\sigma} \pm
c^\dagga_{\bC{\sigma}{\bk}\sigma}\Big)/\sqrt{2}\,,\;
\eal
as well as of hybridisation amplitudes
\be
V_{1(2)\bk\sigma} = \Big(T_{\bk\sigma}\pm T_{\bC{\sigma}{\bk}\sigma}\Big)
/\sqrt{2}\,,
\ee
so that the Hamiltonian can be rewritten as 
\bal
&H(U,\mu,h,V_{2\up},V_{2\down}) = \sum_{\bk\sigma}\,\ep_{\bk\sigma}
\Big(c^\dagger_{1\bk\sigma}\,c^\dagga_{2\bk\sigma}
+ H.c.\Big)\\
& \qquad + \sum_{\bk\sigma}\,\sum_{a=1}^2\,V_{a\bk\sigma}\Big(d^\dagger_\sigma\,c^\dagga_{a\bk\sigma} + 
H.c.\Big)
\\
&\qquad -\fract{U}{4}\,\Omega -\mu\big(n_\up+n_\down-1\big) - h\big(n_\up-n_\down\big)\,,
\label{HAIM-good}
\eal
where we denote the sets of $V_{2\bk\sigma}$ shortly as $V_{2\sigma}$, and hereafter $\bk$ is restricted to $\ep_{\bk\sigma}<0$.\\
Under a spin-$\sigma$ particle-hole transformation 
\be
\mathcal{C}_\sigma\!:\!\Bigg(d^\dagga_\sigma \!\!\to d^\dagger_\sigma\, \cup \,\prod_\bk \bigg(c^\dagga_{1\bk\sigma}\!\!\to\! -c^\dagger_{1\bk\sigma}
\, \cup \, c^\dagga_{2\bk\sigma} \!\!\to c^\dagger_{2\bk\sigma}
\bigg)\Bigg)
\,,\label{C}
\ee
the Hamiltonian parameters change as follows
\bal
& U \to -U\,,\qquad
\mu \to \mp\,h\,,\qquad\;
h \to \mp\,\mu\,,\\
&\qquad\quad\; V_{2\up} \to \mp V_{2\up}\,,\quad
V_{2\down} \to \pm V_{2\down}\,,
\eal
while $V_{1\bk\sigma}$ and $\ep_{\bk\sigma}$ stay invariant. The two signs here refer to the action of $\mathcal{C}_\up$ and 
$\mathcal{C}_\down$, respectively.  
Since the partition function $Z(U,\mu,h,V_{2\up},V_{2\down})$ 
is invariant under any unitary transformation, then  
\bal
Z(U,\mu,h,V_{2\up},V_{2\down}) &= 
Z(-U,-h,-\mu,-V_{2\up},V_{2\down})\\ 
&= Z(-U,h,\mu,V_{2\up},-V_{2\down}) \\
&= Z(U,-\mu,-h,-V_{2\up},-V_{2\down})\,.
\label{ZAIM}
\eal
\section{Mapping within the slave-spin representation}
\label{Mapping within the slave-spin representation}
In Ref.~\onlinecite{De-Medici} a new slave-particle representation of Hubbard-like models was introduced, which in our case consists in associating to each impurity-electron species $d^\dagga_\sigma$ 
an auxiliary Ising variable $\tau^a_\sigma$, $a=x,y,z$. 
The Hamiltonian in such enlarged Hilbert space can be written as
\bal
&H_{2}(U,\mu,h,V_{2\up},V_{2\down}) = \sum_{\bk\sigma}\,\Bigg[\ep_{\bk\sigma}
\Big(c^\dagger_{1\bk\sigma}\,c^\dagga_{2\bk\sigma}
+ H.c.\Big)\\
& \qquad + \tau^x_\sigma\,V_{1\bk\sigma}\Big(d^\dagger_\sigma\,c^\dagga_{1\bk\sigma} + 
H.c.\Big) \\
&\qquad +i\, \tau^y_\sigma\,V_{2\bk\sigma}\Big(d^\dagger_\sigma\,c^\dagga_{2\bk\sigma} - H.c.\Big)\Bigg]
\\
&\qquad +\fract{U}{4}\,\tau^z_\up\,\tau^z_\down 
-\fract{\mu}{2}\big(\tau^z_\up+\tau^z_\down\big) 
- \fract{h}{2}\big(\tau^z_\up-\tau^z_\down\big)\,.
\label{HAIM*-3}
\eal 
This model maps onto the original Hamiltonian Eq.\eqn{HAIM-good} 
in a specified \textit{physical} subspace $\mathscr{H}$ of the enlarged Hilbert space $\mathscr{H}_*$. We introduce the two commuting 
operators
\bal
\mathcal{P}_\up = \tau^z_\up\,\big(2n_\up-1\big)\,,
\qquad \mathcal{P}_\down = \tau^z_\down\,\big(2n_\down-1\big)\,,
\eal
which have eigenvalues $p_\sigma=\pm 1$ and can thus be regarded 
as \textit{parity} operators. The Hamiltonian \eqn{HAIM*-3} commutes with both $\mathcal{P}_\up$ and $\mathcal{P}_\down$, so that each eigenstate of $H_2$ can also be chosen as eigenstate of 
$\mathcal{P}_\sigma$ with eigenvalues $p_\sigma$, $\sigma=\up,\down$. 
The physical subspace $\mathscr{H}$ comprises all states even under parity, i.e. with $p_\sigma=+1$. The projector onto $\mathscr{H}$ is thus
\be
\mathbb{P} = \mathbb{P}_\up\,\mathbb{P}_\down
= \fract{1}{2}\Big(1+\mathcal{P}_\up\Big)\;
\fract{1}{2}\Big(1+\mathcal{P}_\down\Big)\,,\label{P-2}
\ee
and corresponds to the operator equivalence
\be
\tau^z_\sigma \equiv \big(2n_\sigma-1\big)\,,\label{equal}
\ee 
which is just the slave-spin constraint~\cite{De-Medici}. We observe that the hybridisation with 
the operators $c^\dagga_{2\bk\sigma}$ might seem at odds with the original representation 
$d^\dagga_\sigma \to \tau^x_\sigma\,d^\dagga_\sigma$ 
in Ref.~\onlinecite{De-Medici}, but in reality it is not since in the physical subspace $\tau^x_\sigma\,d^\dagger_\sigma \equiv i\tau^y_\sigma\,d^\dagger_\sigma$.
%%%
%since  through Eq.~\eqn{equal} the equalities 
%$d^\dagger_\sigma = \tau^z_\sigma\,d^\dagger_\sigma$ and $d^\dagga_\sigma = -\tau^z_\sigma\,d^\dagga_\sigma$ hold trivially, then
%\ba
%&& -i\, \tau^y_\sigma\,\Big(d^\dagger_\sigma\,c^\dagga_{2\bk\sigma} - H.c.\Big) = -i\, \tau^y_\sigma\,\tau^z_\sigma\,\Big(d^\dagger_\sigma\,c^\dagga_{2\bk\sigma} + H.c.\Big)\nonumber\\
%&&=\tau^x_\sigma\,\Big(d^\dagger_\sigma\,c^\dagga_{2\bk\sigma} + H.c.\Big)\,,
%\ea
%in agreement with the mapping in Ref.~\onlinecite{De-Medici}.  
We shall prefer the expression Eq.~\eqn{HAIM*-3} of the 
slave-spin Hamiltonian, since here the role of the p-h symmetry transformation 
$\mathcal{C}_\sigma$ is simply played by $\tau^x_\sigma$. 
Indeed the equivalences below hold straightforwardly  
\bas
&H_2(-U,h,\mu,V_{2\up},-V_{2\down}) = 
\tau^x_\down\,H_2(U,\mu,h,V_{2\up},V_{2\down})\,\tau^x_\down\,,\\
&H_2(-U,-h,-\mu,-V_{2\up},V_{2\down}) = 
\tau^x_\up\,H_2(U,\mu,h,V_{2\up},V_{2\down})\,\tau^x_\up\,,\\
&H_2(U,-\mu,-h,-V_{2\up},-V_{2\down}) =\\
&\qquad\qquad\qquad\qquad \tau^x_\up\,
\tau^x_\down\,H_2(U,\mu,h,V_{2\up},V_{2\down})\,\tau^x_\down\,
\tau^x_\up,\nonumber
\eas
so that, through Eq.~\eqn{ZAIM}, we find that  
\bal
&Z(U,\mu,h,V_{2\up},V_{2\down}) =
\Tr\bigg(\esp{-\beta H_2(U,\mu,h,V_{2\up},V_{2\down})}\;
\mathbb{P}\bigg)\\
&= \Tr\bigg(\tau^x_\down\,
\esp{-\beta H_1(U,\mu,h,V_{2\up},V_{2\down})}\,\tau^x_\down\,
\mathbb{P}\bigg)\\
&= \Tr\bigg(\tau^x_\up\,
\esp{-\beta H_1(U,\mu,h,V_{2\up},V_{2\down})}\,\tau^x_\up\,
\mathbb{P}\bigg)\\
&= \Tr\bigg(\tau^x_\up\,\tau^x_\down\,
\esp{-\beta H_1(U,\mu,h,V_{2\up},V_{2\down})}\,\tau^x_\down
\,\tau^x_\up\,
\mathbb{P}\bigg)\,.
\eal
Since
$
1=\mathbb{P} + \tau^x_\up\,\mathbb{P}\,\tau^x_\up +\tau^x_\down\,\mathbb{P}\,\tau^x_\down
+\tau^x_\up\,\tau^x_\down\,\mathbb{P}\,
\tau^x_\down\,\tau^x_\up
$, 
it readily follows that 
\be
Z(U,\mu,h,V_{2\up},V_{2\down}) = \fract{1}{4}\,
\Tr\bigg(\esp{-\beta H_2(U,\mu,h,V_{2\up},V_{2\down})}\,\bigg)\,.
\label{final-maping-2}
\ee
The Eq.~\eqn{final-maping-2} is our main result. It states that the partition function of the original impurity model \eqn{HAIM-good} can be calculated \textit{without any constraint} through the partition function of the model \eqn{HAIM*-3}.\\
Following the same line of  reasoning, we can demonstrate that also the physical single-particle Green's functions 
in imaginary time $\tau$ of the impurity can be calculated through the Green's functions of the composite operators $\tau^x_\sigma\,d^\dagga_\sigma$ and $\tau^y_\sigma\,d^\dagga_\sigma$ in the slave-spin representation 
without constraints. In particular (details can be found in the Supplemental Materials) 
\bal 
\mathcal{G}_\sigma(\tau) &= -\big\langle\,\text{T}\Big(
d^\dagga_\sigma(\tau)\,d^\dagger_\sigma(0)\Big)\,\big\rangle\\
&= -\big\langle\,\text{T}\Big(
\tau^x_\sigma(\tau)\,d^\dagga_\sigma(\tau)\;
\tau^+_\sigma(0)\,d^\dagger_\sigma(0)\Big)\,\big\rangle_2\,,\label{Green's functions}
\eal
where $\tau^+_\sigma=\tau^x_\sigma+i\tau^y_\sigma$, and $\langle\dots\rangle_2$ denotes the thermal average with the Boltzmann distribution of 
$H_2$ in  Eq.~\eqn{HAIM*-3} and with the operators propagating in imaginary time with that same Hamiltonian. 

\subsection{An equivalent representation}
The Hamiltonian Eq.~\eqn{HAIM*-3} lacks a clear separation between charge and spin degrees of freedom that is desirable above all when the interaction $U$ is large. The latter is coupled to the combination 
$\tau^z_\up\,\tau^z_\down$, which is therefore the actual operator that controls the large-$U$ freezing of valence fluctuations. Since $\tau^z_\up\,\tau^z_\down$ is still an Ising variable, with value $\pm 1$, we 
can exploit a convenient change of variables and define, following Ref.~\onlinecite{double-frequency}, 
\bal
\tau^z_\up\,\tau^z_\down &= -\sigma^z\,, & \tau^z_\up &= \tau^z\,,&
\tau^z_\down &= -\tau^z\,\sigma^z\,,\\
&&\tau^x_\up &= \tau^x\,\sigma^x\,, &
\tau^x_\down &= \sigma^x\,,\\
&&\tau^y_\up &= \tau^y\,\sigma^x\,, &
\tau^y_\down &= -\tau^z\,\sigma^y\,.\label{new-mapping}
\eal
After this transformation, Eq.~\eqn{HAIM*-3} changes into 
\bea
&&H_{2}(U,\mu,h,V_{2\up},V_{2\down}) = \sum_{\bk\sigma}\,\Bigg[\ep_{\bk\sigma}
\Big(c^\dagger_{1\bk\sigma}\,c^\dagga_{2\bk\sigma}
+ H.c.\Big)\nonumber\\
&&\qquad + \sigma^x\big(\tau^x\,\delta_{\sigma\up}+\delta_{\sigma\down}\big)
\,V_{1\bk\sigma}\Big(d^\dagger_\sigma\,c^\dagga_{1\bk\sigma} + 
H.c.\Big) \nonumber\\
&& +i\, \big(\tau^y\,\sigma^x\,\delta_{\sigma\up}-\tau^z\,\sigma^y\,\delta_{\sigma\down}\big)
\,V_{2\bk\sigma}\Big(d^\dagger_\sigma\,c^\dagga_{2\bk\sigma} - H.c.\Big)\Bigg]\nonumber\\
&&-\fract{U}{4}\,\sigma^z
-\bigg[\fract{\mu}{2}\big(1-\sigma^z\big) 
+\fract{h}{2}\big(1+\sigma^z\big)\bigg]\,\tau^z\,,\label{HAIM*-4}
\eea 
where $\delta_{\sigma\sigma'}$ is the Kronecker delta. 
Eq.~\eqn{HAIM*-4} notably simplifies when $V_{2\sigma}=0$. In this case 
$\mathcal{P}_\up = \tau^z\,\big(2n_\up-1\big)$, with eigenvalues $p_\up=\pm 1$, is conserved, and moreover the two subspaces with $p_\up=\pm1$ are actually related by the p-h transformation $\mathcal{C}_\up$ Eq.~\eqn{C}. Therefore, following exactly the same steps as before but in the reverse order, we conclude that the partition function of the original model Eq.~\eqn{HAIM-good} at $V_{2\sigma}=0$ can be calculated as 
\bal
Z(U,\mu,h,0,0) &= \fract{1}{2}\,\Tr\bigg(\esp{-\beta H_1(U,\mu,h)}\,\bigg)\,,\label{mapping-ph}
\eal
where 
\bea
&&H_{1}(U,\mu,h) = \sum_{\bk\sigma}\,\Bigg[\ep_{\bk\sigma}
\Big(c^\dagger_{1\bk\sigma}\,c^\dagga_{2\bk\sigma}
+ H.c.\Big) \nonumber\\
&&\qquad + \sigma^x
\,V_{1\bk\sigma}\Big(d^\dagger_\sigma\,c^\dagga_{1\bk\sigma} + 
H.c.\Big)\Bigg] \label{HAIM*-5}\\
&&
-\fract{U}{4}\,\sigma^z
-\bigg[\fract{\mu}{2}\big(1-\sigma^z\big) 
+\fract{h}{2}\big(1+\sigma^z\big)\bigg]\,\big(2n_\up-1\big)\,,\nonumber
\eea
involves now a single auxiliary Ising variable. The mapping Eq.~\eqn{mapping-ph} with the Hamiltonian \eqn{HAIM*-5} generalises the results obtained in Ref.~\onlinecite{Marco-PRB} in the presence of a chemical shift of the impurity level, both spin independent and dependent. 

\subsection{Extension to multi-orbital impurity models}
The mapping in Sec.~\ref{Mapping within the slave-spin representation} can be straightforwardly extended to a multi-orbital impurity model with Hamiltonian    
\bal
H &= H_\text{imp} + \sum_{\bk\sigma}\,\sum_{\alpha=1}^M\,
\ep_{\alpha\bk\sigma}\,\Big(c^\dagger_{1\alpha\bk\sigma}\,c^\dagga_{2\alpha\bk\sigma}
+ H.c.\Big)\\
&  + \sum_{\bk\sigma}\,\sum_{a=1}^2\,
\sum_{\alpha=1}^M\,V_{a \alpha \bk\sigma}\Big(d^\dagger_{\alpha\sigma}\,c^\dagga_{a \alpha \bk\sigma}+ 
H.c.\Big)\,,\label{HAIM-multi}
\eal
in the simple and not very realistic case where the isolated impurity Hamiltonian 
$H_\text{imp}$ involves only the occupation numbers 
$n_{\alpha\sigma}=d^\dagger_{\alpha\sigma}\,d^\dagga_{\alpha\sigma}$, where $\alpha=1,\dots,M$ is the orbital index, i.e. 
$H_\text{imp} = H_\text{imp}\Big(\{n_{\alpha\sigma}\}\Big)$, 
does not include Coulomb exchange terms. 
In this circumstance we can exploit the p-h transformations Eq.~\eqn{C} for each orbital species and follows exactly the same reasoning as in 
Sec.~\ref{Mapping within the slave-spin representation} to show that the partition function $Z$ of the Hamiltonian \eqn{HAIM-multi} 
can be calculated through 
\be
Z = \left(\fract{1}{2}\right)^{2M}\,
\Tr\bigg(\text{e}^{-\beta H_*}\bigg)\,,
\ee
where 
\bea
H_* &=& H_\text{imp}\big(\big\{\tau^z_{\alpha\sigma}\big\}\big) + 
\sum_{\alpha \bk\sigma}\,
\ep_{\bk \alpha\sigma}\Big(c^\dagger_{1\alpha \bk \sigma}\,c^\dagga_{2\alpha \bk \sigma}
+ H.c.\Big)\nonumber\\
&&  + \sum_{\alpha \bk\sigma}\,\bigg[\tau^x_{a\sigma}\,V_{1\bk a\sigma}\,
\Big(d^\dagger_{a\sigma}\,
c^\dagga_{1\bk a\sigma} + 
H.c.\Big)\nonumber\\
&& \qquad\qquad + i\tau^y_{a\sigma}\,V_{2\bk a\sigma}\,
\Big(d^\dagger_{a\sigma}\,c^\dagga_{2\bk a\sigma} - H.c.\Big)\bigg]
\,.\label{HAIM-multi-*}
\eea

%%%%%%%%%%
\section{Mean field solution}
\label{Mean field solution in the presence of a Zeeman splitting}
To highlight the importance of a mapping without constraints, we here study the simple case where 
the bath and the hybridisation are both p-h invariant and the only source of p-h asymmetry is either a Zeeman splitting $h$ or a chemical shift $\mu$ of the impurity level. The Hamiltonian is therefore  
that in Eq.~\eqn{HAIM-good} at finite $h\gtrsim 0$ but $\mu=0$,  or vice versa, 
with $V_{2\bk\sigma} = 0$ and spin-independent $\ep_{\bk\sigma} = \ep_\bk$ and $V_{1\bk\sigma}=V_\bk$.\\
We mention that the mean-field approach to the standard slave-boson representation of such Hamiltonian 
at $h=\mu=0$ erroneously yields at large $U$ a 
negative magnetic susceptibility $\chi_\text{imp}<0$, see Supplemental Materials for details, signalling instability of the paramagnetic solution towards spontaneous spin polarisation~\cite{Schonhammer}. This is the tangible evidence that imposing the constraint \textit{on average} may lead to wrong results.\\
Let us consider instead our mapping onto the equivalent Hamiltonians \eqn{HAIM*-4} and \eqn{HAIM*-5}, which do not require any constraint to be imposed. The simplest mean-field approach consists in approximating the ground state wavefunction with a factorised one product of a fermionic part $\ket{\!\Psi}$ times an Ising one $\ket{\!\Phi}$. However, such an approximation is physically sound as long as the two subsystems are controlled by well separated energy scales, otherwise we have no guarantee that the fluctuations beyond mean field are negligible. This is indeed realised in model \eqn{HAIM*-5}
when $U$ is large.  On the contrary, a sharp distinction of energy scales is 
absent in the equivalent representation Eq.~\eqn{HAIM*-4}, where, after 
mean-field decoupling, the Ising sector $\big(\boldsymbol{\tau},\boldsymbol{\sigma}\big)
\equiv \big(\boldsymbol{\tau}_\up,\boldsymbol{\tau}_\down\big)$ always contains excitation energies 
within the resonant level spectral width. Therefore, even though Eq.~\eqn{HAIM*-4} is equivalent to Eq.~\eqn{HAIM*-5}, the mean-field approximation is only justified in the latter model and when $U$ is large,
which we shall consider hereafter. \\    
Within mean-field applied to model \eqn{HAIM*-4}, if we denote as
\be
\sin\theta = \bra{\Phi}\sigma^x\ket{\Phi}\,,\qquad \cos\theta = \bra{\Phi}\sigma^z\ket{\Phi}\,,
\label{theta}
\ee
then the optimal $\ket{\Psi}$ is the ground state of the Hamiltonian 
\ba
H_* &=& \sum_{\bk\sigma}\,\Bigg[\ep_{\bk\sigma}\,c^\dagger_{\bk\sigma}\,c^\dagga_{\bk\sigma} 
+ \sin\theta\,V_{\bk}\Big(d^\dagger_\sigma\,c^\dagga_{\bk\sigma} + c^\dagger_{\bk\sigma}\,d^\dagga_\sigma\Big)\Bigg] \\
&&\qquad -\ep\,\big(1\pm\cos\theta\big)\left(n_\up-\fract{1}{2}\right)\,,
\ea
where the plus sign applies to $\ep=h$, while the minus to $\ep=\mu$. 
Assuming, as usual, that the hybridisation function $\Delta(\omega)$ with the bath can be approximated as 
\be
\Delta(\omega)=\sum_\bk\,\fract{V_\bk^2}{\omega-\ep_\bk+i0^+} \simeq -i\,\Gamma_0\,\theta\big(D-|\omega|\big)\,,
\ee
where the cut-off $D$ is of the order of the conduction bandwidth, we readily find that 
\bal
E_*(\theta) &\equiv \bra{\Psi} H_* \ket{\Psi} = E_0 - \ep_\up(\theta)\;\left(n_\up(\theta)-\fract{1}{2}\right)\\
& -\frac{\Gamma(\theta)}{\pi}\,\Bigg[
\ln \fract{\text{e}D}{\Gamma(\theta)} 
+ \ln \fract{\text{e}D}{\sqrt{\ep_\up(\theta)^2+\Gamma(\theta)^2}} 
\Bigg]\,,
\eal
where $E_0$ is the bath-energy in the absence of impurity, $\Gamma(\theta) =\sin^2\theta\,\Gamma_0$
and 
\[
\ep_\up(\theta)  = \ep\,\big(1\pm\cos\theta\big)\,, \quad n_\up(\theta)-\fract{1}{2} = \fract{1}{\pi}\,\tan^{-1} \fract{\ep_\up(\theta)}{\Gamma(\theta)}\,.
\]
The variational energy is therefore 
\bal
E(\theta) &= \bra{\Phi}\!\bra{\Psi} H_1(U,0,h) \ket{\Psi}\!\ket{\Phi}= E_*(\theta) - \fract{U}{4}\,\cos\theta\,,\nonumber
\eal
which we still have to minimise with respect to $\theta$. 
It is more convenient to use $\Gamma = \Gamma\big(\theta(\Gamma)\big)$ as variational parameter, which leads to the saddle-point equation 
\bal
0 &=\fract{\partial E(\Gamma)}{\partial \Gamma} =
-\frac{1}{\pi}\,\Bigg[
\ln \fract{D}{\Gamma} 
+ \ln \fract{D}{\sqrt{\ep_\up\big(\theta(\Gamma)\big)^2+\Gamma^2}}\Bigg] 
\\
&\; +\Bigg(\fract{U}{4}\pm\fract{\ep}{\pi}\,\tan^{-1} \fract{\ep_\up\big(\theta(\Gamma)\big)}{\Gamma}\Bigg)\,\fract{1}{2}\;\fract{1}{\sqrt{\Gamma_0^2 
-\Gamma_0\,\Gamma\;}}\,.\label{saddle-point}
\eal
For large $U$ the solution of  Eq.~\eqn{saddle-point} at $\ep\ll \Gamma$ reads
\bal
\Gamma(\ep) &\simeq \Gamma(0) - \fract{\ep^2}{4\Gamma(0)}\,\Big(1\pm\sqrt{1-\Gamma(0)/\Gamma_0\,}\;\Big)^2 \,,\label{T_K}
\eal
where $\Gamma(0)\simeq D\,\exp\big[-\pi U/16\Gamma_0\big]$ is the same as in slave-boson mean-field theory, and can be associated with the Kondo temperature $T_K$, though overestimated with respect to its actual value~\cite{Pierpaolo}. The susceptibility to the field $\ep$ readily follows 
\be
-\fract{\partial^2 E}{\partial \ep^2}_{\big|\ep=0} \simeq \fract{1}{\pi\Gamma(0)}\,\Big(1\pm\sqrt{1-\Gamma(0)/\Gamma_0\,}\;\Big)^2
\,.\label{chi-epsilon}
\ee
Since $\Gamma_0\gg \Gamma(0)$ for $U\gg \Gamma_0$, the impurity contribution 
to charge $\kappa_\text{imp}$, $\ep=\mu$ and minus sign, and spin $\chi_\text{imp}$, $\ep=h$ and plus sign, susceptibilities become  
\bal
\kappa_\text{imp} &\simeq
\fract{\Gamma(0)}{4\pi\Gamma_0^2}\simeq 0\,,\\
\chi_\text{imp} &\simeq \fract{4}{\pi\Gamma(0)}\,\left(1-\fract{\Gamma(0)}{2\Gamma_0}\right)\simeq 
\fract{4}{\pi\Gamma(0)}
\,.\label{chi&kappa}
\eal
We emphasize that $\chi_\text{imp}$ is positive, unlike in slave-boson mean-field theory. The impurity contribution to the specific heat at low temperature only comes from the fermionic degrees of freedom and reads explicitly 
\be 
c_{\text{imp}} \simeq \fract{2\pi^2}{3}\, \fract{T}{\pi\Gamma(0)}\,,\label{c_v}
\ee
thus a Wilson ratio $R_\textit{W}=2$ at large $U$, in agreement with the exact value.\\
According to Nozi\`eres' Fermi liquid description of the Kondo effect~\cite{Nozieres_FL}, see also Ref.~\onlinecite{Fred1978}, 
\be
\kappa_\text{imp} = 2\rho_*\,\Big(1-A^\text{S}\Big)\,,\qquad 
\chi_\text{imp} = 2\rho_*\,\Big(1-A^\text{A}\Big)\,,\label{FL-susceptibility}
\ee
where $\rho_* = 1/\pi\Gamma(0) = Z\rho_0$ is the quasiparticle density of states at the chemical potential, 
as opposed to its \textit{bare} value $\rho_0=1/\pi\Gamma_0$, with $Z=\sin^2\theta\ll 1$ the quasiparticle residue;  while  $A^\text{S}$ and $A^\text{A}$ the quasiparticle scattering amplitudes in the symmetric (S) and antisymmetric (A) channels, respectively. The mean-field results \eqn{chi&kappa} are thus compatible at large $U$ with 
\be
A^\text{S} = - A^\text{A} = 1\,,\label{A-FL}
\ee
which, together with Eq.~\eqn{c_v}, are the bases of Nozi\`eres' local Fermi liquid theory of the Kondo effect~\cite{Nozieres_FL},  which has been 
successfully exploited in very many contexts, not least to derive universal properties in transport across quantum dots~\cite{SashaPRL2006,Referee}. We emphasise that the universal values in Eq.~\eqn{A-FL} simply follows from the expressions of the impurity charge and spin density 
vertices, the former proportional to $(1-\sigma^z)$ and the latter to $(1+\sigma^z)$, and the fact that, at large 
$U$, $\sigma^z\simeq 1$ with negligible fluctuations. As a result, the mean-field solution, 
$\sigma^z\to \langle\sigma^z\rangle$, already captures the leading vertex corrections, which is indeed  remarkable. 
%%%%%%%%  ADDED TEXT
By contrast, the mean-field approximation does not allow recovering the non-universal 
corrections to the Kondo regime, which are polynomials in $1/U$ for large $U$~\cite{Velko1985,Evers2012}. 
These corrections are sub-leading in the spin susceptibility, but leading in the charge one, see 
Eq.~\eqn{chi&kappa}.      
%%%%%%%% END ADDED TEXT   
\\
%%%%%%%%%%%%%%% FIG
\begin{figure}[thb]
\vspace{-0.2cm}
\centerline{\includegraphics[width=6.3cm,angle=-90]{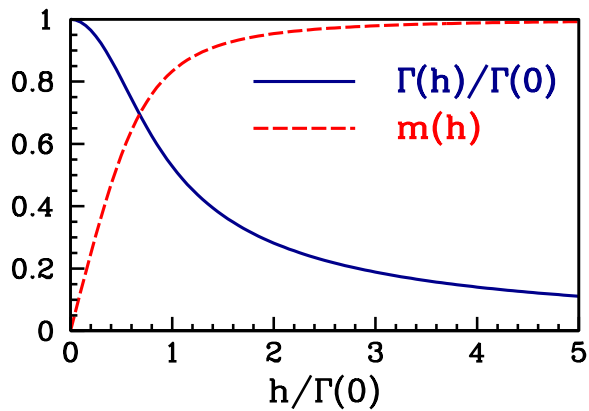}}
\vspace{-0.3cm}
\caption{Mean field values of the effective hybridisation width $\Gamma(h)$ and magnetisation $m(h)$ as function of the magnetic field $h$. The parameters are $U=0.1$ and $\Gamma_0=1.96\times10^{-3}$ in units of the cutoff 
$D$, which correspond to $\Gamma(0)\equiv T_K \simeq 4.1\times 10^{-5}$.} 
\label{TK.vs.h}
\end{figure}
%%%%%%%%%%%%%%%%%
We conclude by mentioning that the model Eq.~\eqn{HAIM*-5} can be still viewed as a dissipative two-level system~\cite{Caldeira&Leggett} in a sub-ohmic bath, as it was the case at $\ep=0$~\cite{Pierpaolo}. Each potential well corresponds to a value of 
$\sigma^x=\pm 1$, while $\sigma^z$ induces quantum tunnelling between the two wells. Localisation inside a well is signalled by a finite expectation value of $\sigma^x$, and also corresponds to spontaneous breakdown of the local $Z_2$ gauge symmetry $\sigma^x\to-\sigma^x$ and $d^\dagga_\sigma \to - d^\dagga_\sigma$. The Kondo temperature $T_K\sim \langle\sigma^x\rangle^2$ thus plays the role of a \textit{bona fide} order parameter.\\  
In this language, the field $\ep$ translates into an assisted tunnelling that does hamper localisation but, at least within mean-field, cannot impede it, as shown in Fig.~\ref{TK.vs.h} for the case of a Zeeman splitting $\ep=h$, where we plot the mean field values of  $\Gamma(h)\sim \langle\sigma^x\rangle_h^2$ and magnetisation $m(h)$. We believe that the 
persistence of $Z_2$ gauge-symmetry breaking even in the presence of the assisted tunnelling is real and not just an artefact of mean 
field.\\
    
\section{Conclusions}
We have shown that a generic single-orbital Anderson impurity model can be mapped without any constraint onto a resonant level model coupled to two Ising spins, or just one in the simpler case when the hybridisation with the bath is particle-hole symmetric. The mean-field decoupling of electrons from the Ising variables is able to reproduce quite accurately the magnetic properties of the model even deep inside the large-$U$ Kondo regime, specifically the finite susceptibility $\chi\sim 1/T_K$ and Wilson ratio $R_\text{W}=2$.
By comparison, in the same Kondo regime conventional slave-boson mean-field theory yields a spin-polarised lowest energy solution that unphysically breaks spin $SU(2)$ symmetry.\\
We also demonstrate how single-particle Green's functions of the physical fermions can be calculated without constraints, which would for instance allow exploiting DMFT to study in the slave-spin 
representation~\cite{Rok-PRB2015} particle-hole non-symmetric Hubbard-like models in lattices with infinite coordination. This could in some cases be more convenient than directly working within the physical Hilbert space, though smaller, especially when one wants to prevent spontaneous symmetry breaking that usually accompanies a Mott transition, since the slave-spin Hamiltonian Eq.~\eqn{HAIM*-3} is particle-hole symmetric in terms of the auxiliary fermions, despite the Hamiltonian of the physical electrons is not. 

\section*{Acknowledgments}
This work has been supported by 
the European Union under H2020 Framework Programs, ERC Advanced Grant No. 692670 ``FIRSTORM''.

%%%%%%%%%%%%%%% FIG
%\begin{figure}[t]
%\vspace{-0.8cm}
%\centerline{\includegraphics[width=6.3cm,angle=-90]{*.eps}}
%\vspace{-1cm}
%\caption{} 
%\label{}
%\end{figure}
%%%%%%%%%%%%%%%%%

%\section*{Acknowledgments}
%This work has been supported by 
%the European Union, Seventh Framework Programme, under the project
%GO FAST, grant agreement no. 280555. 
    
\bibliographystyle{apsrev}
%\bibliography{mybiblio}

\end{document}

\centerline{\Huge\textbf{Supplemental materials}}
\bigskip
\bigskip
\bigskip
\section{Slave-boson mean field approximation in a magnetic field}
The mean-field approximation within the paramagnetic sector of the slave-boson representation of an Anderson impurity model is long since known~\cite{Barnes-slave-boson,Coleman-slave-boson,Read&Newns,Read-quantum-fluctuations,Bickers-RMP}. However the mean-field results allowing for spontaneous magnetisation of the impurity are not as widely known. It was mentioned in Ref.~\cite{Schonhammer} that at large $U$ the actual lowest-energy mean-field solution is magnetic, though no details were presented. For this reason we think it is worth to give here all details of such calculation. \\
It is known that the slave-boson mean-field theory in the consistent formulation of Kotliar and Ruckenstein~\cite{Kotliar&Ruckenstein} is equivalent to the Gutzwiller 
approximation~\cite{Gebhard-PRB1991},  so we shall use the latter technique, for which we refer to 
Ref.~\cite{Hvar} for details. \\    
Within the Gutzwiller approximation the variational energy reads
\bal
E(\theta,m) = E_*(\theta,m) - \fract{U}{4}\,\cos\theta - h\,m\,,
\eal
where $E_*(\theta,m)$ is the lowest expectation value of the resonant level Hamiltonian 
\be
H_* \!= \!\sum_{\bk\sigma}\!\bigg[\ep_{\bk\sigma}\,c^\dagger_{\bk\sigma}\,c^\dagga_{\bk\sigma} 
+ R(\theta,m)\,V_{\bk}\Big(d^\dagger_\sigma\,c^\dagga_{\bk\sigma} + c^\dagger_{\bk\sigma}\,d^\dagga_\sigma\Big)\bigg]\,,\label{H-G}
\ee
within a subspace of wavefunctions $\ket{\Psi}$ such that  $\bra{\Psi} n_\up-n_\down \ket{\Psi}=m$. 
The hybridisation is renormalised downwards by the quantity
\bal
R(\theta,m) &= \fract{1}{\sqrt{1-m^2}}\;\sin\fract{\theta}{2}\;
\Bigg[\sqrt{ \cos^2\fract{\theta}{2} +m\;} + \sqrt{ \cos^2\fract{\theta}{2} -m\;}\;\Bigg]\\
&\simeq \sin\theta\,\Bigg[1+ \fract{1}{2}\,m^2\,\fract{\cos\theta\,\big(2+\cos\theta\big)}{\big(1+\cos\theta\big)^2}\Bigg]\\
&\equiv \sin\theta \Big(1+ \fract{\rho(\theta)}{2}\,m^2\Big)\,
\eal
where the second expression is the expansion for $\cos^2\theta/2 \gg m$. We define the effective hybridisation width 
\be
\Gamma(\theta,m) = R(\theta,m)^2\,\Gamma_0\simeq \Gamma(\theta)\Big(1+\rho(\theta)\,m^2\Big)\,,
\ee
with $\Gamma(\theta)=\Gamma_0\,\sin^2\theta$, being $\Gamma_0$ its unrenormalized value, 
and an effective field $h_*(\theta,m)$ such that the magnetisation has the desired value, which corresponds to 
the solution of the following equation 
\be
m = \fract{2}{\pi}\,\tan^{-1} \fract{h_*(\theta,m)}{\Gamma(\theta,m)}\,,
\ee
which, at small $m$, is simply 
\be
h_*(\theta,m)\simeq \fract{\pi\,\Gamma(\theta)}{2}\, m
\ee
\begin{figure}[t]
\centerline{\includegraphics[width=0.8\textwidth]{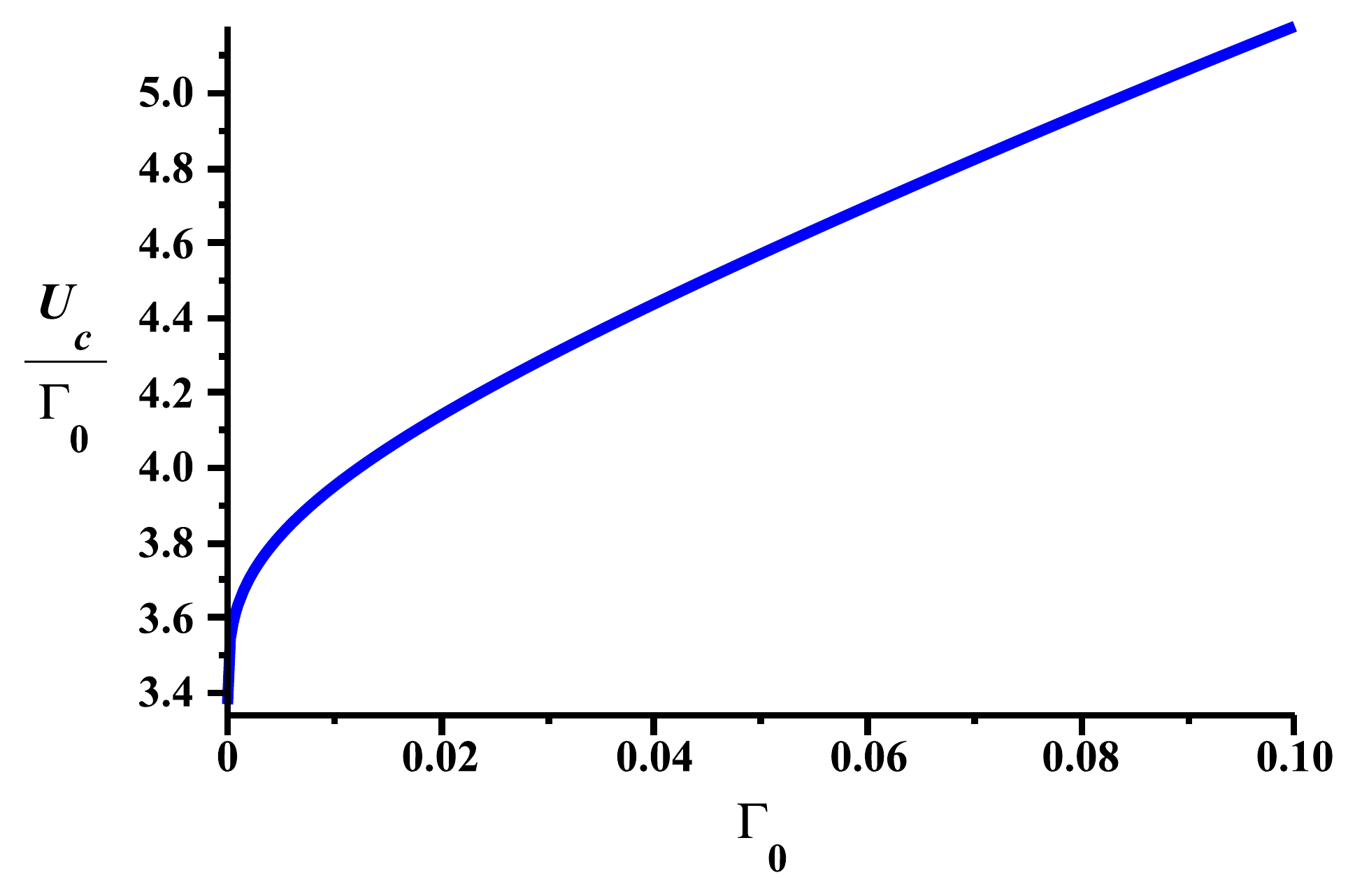}}
\caption{Critical $U_c$ above which the lowest energy mean-field solution is magnetic, in units of the bare hybridisation width $\Gamma_0$. The energy unit is the cutoff $D=1$.}
\label{fig-1}
\end{figure}
With those definition the variational energy is readily found to be
\bal
E(\theta,m) &= E_0 -2\,\frac{\Gamma(\theta,m)}{\pi}\,
\ln \fract{\text{e}D}{\sqrt{h_*(\theta,m)^2+\Gamma(\theta,m)^2\;}} \\
&  - \fract{U}{4}\,\cos\theta - h\,m\,,
\eal
and has to be minimised with respect to $\theta$ and $m$. Let us 
first study at $h=0$ the stability of the $m=0$ solution towards developing a spontaneous magnetisation $m\ll 1$. 
Expanding the energy we find
\bal
E(\theta,m) &\simeq E_0  - 2\,\frac{\Gamma(\theta)}{\pi}\;\Big(1+\rho(\theta)\,m^2\Big)\,\Bigg[
\ln \fract{\text{e}D}{\Gamma(\theta)\;}
-\fract{1}{2}\,\ln\bigg(1+2\rho(\theta)\,m^2 + \fract{\pi^2}{4}\,m^2\bigg)\Bigg]- \fract{U}{4}\,\cos\theta\\
&\simeq E_0  - 2\,\frac{\Gamma(\theta)}{\pi}\;
\ln \fract{\text{e}D}{\Gamma(\theta)\;}- \fract{U}{4}\,\cos\theta+ 2\,\frac{\Gamma(\theta)}{\pi}\;m^2\,\Bigg\{
\fract{\pi^2}{8} - \rho(\theta)\,\ln \fract{D}{\Gamma(\theta)\;}
\Bigg\}\,.
\eal 
The paramagnetic solution is stable as long as the expression $\Xi(\theta)$ in the curly bracket is positive at the saddle point value $\theta = \theta_*$ with $m=0$, which satisfies
\bal
\fract{4\Gamma_0}{\pi}\,\cos\theta_*\;\ln\fract{D}{\;\Gamma(\theta_*)\;} 
=\fract{U}{4}\;. 
\eal
It follows that 
\bal
\Xi(\theta_*) &= \fract{\pi^2}{8} - \rho(\theta_*)\;\fract{\pi U}{16\Gamma_0\,\cos\theta_*}
>0, 
\eal
is the stability condition, while the equality define the critical $U_c$ above which the 
lowest energy solution is magnetic, shown in Fig.~\ref{fig-1}.  For instance, when $\Gamma_0\ll D$ we find $U_c\simeq 3.24\,\Gamma_0$,  so that in the Kondo regime $U\gg \Gamma_0$ the lowest energy solution is magnetic, which is evidently unphysical.    

\section{Single-particle Green's functions in the physical subspace}
Following similar arguments that allow calculating the physical partition function through the partition function of the slave-spin model without constraints, one can show that also the impurity single-particle Green's functions in the physical subspace can be calculated through appropriate correlation functions in the full Hilbert space with no constraint.  \\

In the original representation, the impurity single-particle Green's functions in imaginary time are defined through 
\bal
\mathcal{G}_\sigma(\tau) = 
-\langle \text{T}\Big(d^\dagga_\sigma(\tau)\,d^\dagger_\sigma(0)\Big)
\rangle &= 
-\fract{\theta(\tau)}{Z}\,\Tr\bigg(
\text{e}^{-\beta H}\;\text{e}^{\tau H}\;
d^\dagga_\sigma\,\text{e}^{-\tau H}\;d^\dagger_\sigma\bigg)\\
&\; + \fract{\theta(\tau)}{Z}\,\Tr\bigg(
\text{e}^{-\beta H}\;d^\dagger_\sigma\,\text{e}^{\tau H}\;
d^\dagga_\sigma\,\text{e}^{-\tau H}\;\bigg)\,,
\eal
where the Hamiltonian $H=H(U,\mu,h,V_{2\up},V_{2\down})$. 
Through the action of the particle-hole transformations 
$\mathcal{C}_\up$ and $\mathcal{C}_\down$, and exploiting the invariance of the trace under a unitary transformation, one readily finds the following relationships between the Green's functions in Matsubara frequencies $i\ep$:
\bal
\mathcal{G}_\up(i\ep;U,\mu,h,V_{2\up},V_{2\down}) &= 
-\mathcal{G}_\up(-i\ep;-U,-h,-\mu,-V_{2\up},V_{2\down}) 
= \mathcal{G}_\up(i\ep;-U,h,\mu,V_{2\up},-V_{2\down})\\
&= -\mathcal{G}_\up(-i\ep;U,-\mu,-h,-V_{2\up},-V_{2\down})\\
\mathcal{G}_\down(i\ep;U,\mu,h,V_{2\up},V_{2\down}) &= 
\mathcal{G}_\down(i\ep;-U,-h,-\mu,-V_{2\up},V_{2\down}) 
= -\mathcal{G}_\down(-i\ep;-U,h,\mu,V_{2\up},-V_{2\down})\\
&= -\mathcal{G}_\down(-i\ep;U,-\mu,-h,-V_{2\up},-V_{2\down})\,.
\label{Green's}
\eal  
Because of Eq.~\eqn{Green's}, we can therefore define the physical Green's functions as
\bal
\mathcal{G}_\up(i\ep;U,\mu,h,V_{2\up},V_{2\down}) &\equiv
\fract{1}{2}\,\bigg(\mathcal{G}_\up(i\ep;U,\mu,h,V_{2\up},V_{2\down})
+ \mathcal{G}_\up(i\ep;-U,h,\mu,V_{2\up},-V_{2\down})\bigg)\,,\\
\mathcal{G}_\down(i\ep;U,\mu,h,V_{2\up},V_{2\down}) &= 
\fract{1}{2}\,\bigg(\mathcal{G}_\down(i\ep;U,\mu,h,V_{2\up},V_{2\down})
+ \mathcal{G}_\down(i\ep;-U,-h,-\mu,-V_{2\up},V_{2\down})\bigg)\,.
\label{Green's-2}
\eal  
In the slave spin representation, the physical Green's functions can be alternatively obtained using the 
slave-spin Hamiltonian 
\bal
&H_{2}(U,\mu,h,V_{2\up},V_{2\down}) = \sum_{\bk\sigma}\,\Bigg[\ep_{\bk\sigma}
\Big(c^\dagger_{1\bk\sigma}\,c^\dagga_{2\bk\sigma}
+ H.c.\Big) + \tau^x_\sigma\,V_{1\bk\sigma}\Big(d^\dagger_\sigma\,c^\dagga_{1\bk\sigma} + 
H.c.\Big)\\
&\qquad\qquad \qquad \qquad \qquad \qquad \qquad   -i\, \tau^y_\sigma\,V_{2\bk\sigma}\Big(d^\dagger_\sigma\,c^\dagga_{2\bk\sigma} - H.c.\Big)\Bigg]
\\
&\qquad +\fract{U}{4}\,\tau^z_\up\,\tau^z_\down 
-\fract{\mu}{2}\big(\tau^z_\up+\tau^z_\down\big) 
- \fract{h}{2}\big(\tau^z_\up-\tau^z_\down\big)\,,
\label{HAIM*-3}
\eal 
through the following expressions:
\bal
\mathcal{G}_\sigma(\tau) &= 
-\fract{\theta(\tau)}{Z}\,\Tr\bigg(
\text{e}^{-\beta H_2}\;\text{e}^{\tau H_2}\;\tau^x_\sigma\,
d^\dagga_\sigma\,\text{e}^{-\tau H_2}\;\tau^x_\sigma\,d^\dagger_\sigma\;
\mathbb{P}_\up\,\mathbb{P}_\down\bigg)\\
&\; + \fract{\theta(\tau)}{Z}\,\Tr\bigg(
\text{e}^{-\beta H_2}\;\tau^x_\sigma\,d^\dagger_\sigma\,\text{e}^{\tau H_2}\;\tau^x_\sigma\,
d^\dagga_\sigma\,\text{e}^{-\tau H_2}\;\mathbb{P}_\up\,\mathbb{P}_\down\bigg)\,,
\eal
where 
\be
\mathbb{P}_\sigma = \fract{1}{2}\,\bigg(1+\tau^z_\sigma\,
\big(2n_\sigma-1\big)\bigg)\,, 
\ee 
is a projector and, by definition, 
\be
\overline{\mathbb{P}}_\sigma = 1- \mathbb{P}_\sigma 
= \mathcal{C}_\sigma\, \mathbb{P}_\sigma\,\mathcal{C}_\sigma
= \tau^x_\sigma\,\mathbb{P}_\sigma\,\tau^x_\sigma\,.
\ee
Noting that, e.g., 
\bal
&\Tr\bigg(
\text{e}^{-\beta\, \mathcal{C}_\down\,H\,\mathcal{C}_\down}\;\text{e}^{\tau \,\mathcal{C}_\down\,H\,\mathcal{C}_\down}\;
d^\dagga_\up\,\text{e}^{-\tau \,\mathcal{C}_\down\,H\,\mathcal{C}_\down}\;d^\dagger_\up\bigg)\\
&= \Tr\bigg(
\text{e}^{-\beta\, \tau^x_\down\,H_2\,\tau^x_\down}\;\text{e}^{\tau \,\tau^x_\down\,H_2\,\tau^x_\down}\;\tau^x_\up\,
d^\dagga_\up\,\text{e}^{-\tau \,\tau^x_\down\,H_2\,\tau^x_\down}\;\tau^x_\up\,d^\dagger_\up\;\mathbb{P}_\up\,\mathbb{P}_\down\bigg)\\
&= \Tr\bigg(
\text{e}^{-\beta\, H_2}\;\text{e}^{\tau \,H_2}\;\tau^x_\up\,
d^\dagga_\up\,\text{e}^{-\tau \,H_2}\;\tau^x_\up\,d^\dagger_\up\;\mathbb{P}_\up\,\overline{\mathbb{P}}_\down\bigg)\,,
\eal
if we instead use the alternative definition in Eq.~\eqn{Green's-2}, 
then the Green's functions can be equivalently calculated through 
\bal
\mathcal{G}_\sigma(\tau) &= 
-\fract{\theta(\tau)}{2Z}\,\Tr\bigg(
\text{e}^{-\beta H_2}\;\text{e}^{\tau H_2}\;\tau^x_\sigma\,
d^\dagga_\sigma\,\text{e}^{-\tau H_2}\;\tau^x_\sigma\,d^\dagger_\sigma\;
\mathbb{P}_\sigma\bigg)\\
&\; + \fract{\theta(\tau)}{2Z}\,\Tr\bigg(
\text{e}^{-\beta H_2}\;\tau^x_\sigma\,d^\dagger_\sigma\,\text{e}^{\tau H_2}\;\tau^x_\sigma\,
d^\dagga_\sigma\,\text{e}^{-\tau H_2}\;\mathbb{P}_\sigma\bigg)\,,
\label{Green's-3}
\eal
where $Z$ is still the partition function of the physical system. We then observe that 
\bal
\tau^+_\sigma\,d^\dagger_\sigma &= \Big(\tau^x_\sigma +
i\tau^y_\sigma\Big)\,d^\dagger_\sigma = 2\tau^x_\sigma\,d^\dagger_\sigma\,\mathbb{P}_\sigma\,,
\eal
so that Eq.~\eqn{Green's-3} can be also written as 
\bal
\mathcal{G}_\sigma(\tau) &= 
-\fract{\theta(\tau)}{4Z}\,\Tr\bigg(
\text{e}^{-\beta H_2}\;\text{e}^{\tau H_2}\;\tau^x_\sigma\,
d^\dagga_\sigma\,\text{e}^{-\tau H_2}\;\tau^+_\sigma\,d^\dagger_\sigma\bigg)\\
&\; + \fract{\theta(\tau)}{4Z}\,\Tr\bigg(
\text{e}^{-\beta H_2}\;\tau^+_\sigma\,d^\dagger_\sigma\,\text{e}^{\tau H_2}\;\tau^x_\sigma\,
d^\dagga_\sigma\,\text{e}^{-\tau H_2}\bigg)\\
&= -\, \big\langle \,\text{T}\Big(\tau^x_\sigma(\tau)\,
d^\dagga_\sigma(\tau)\;\tau^+_\sigma(0)\,d^\dagger_\sigma(0)\Big)\,\big\rangle_2 \,,
\label{Green's-4}
\eal
where the thermal average 
\be
\langle \dots\rangle_2 \equiv 
\fract{\Tr\Big(\text{e}^{-\beta\,H_2}\;\dots\Big)}
{\Tr\Big(\text{e}^{-\beta\,H_2}\,\Big)}\;,
\ee
is performed without constraints. The equation \eqn{Green's-4} thus allows calculating the single-particle Green's functions of the physical particles in terms of the Green's functions of the composite operators 
$\tau^x_\sigma\,d^\dagga_\sigma$ and $\tau^y_\sigma\,d^\dagga_\sigma$ in the slave-spin representation 
without constraints. This result is not only useful by its own but also because it allows implementing 
DMFT~\cite{DMFT,Rok-PRB2015} in the slave-spin representation of Hubbard-like models in lattices with infinite coordination number, which in some cases could be more convenient than working directly in the physical representation. 

%\begin{figure}[t]
%\centerline{\includegraphics[width=7cm]{*.pdf}}
%\caption{}
%\label{fig-1}
%\end{figure}

\bibliographystyle{unsrt}
%\bibliography{mybiblio}